\documentclass[%
 aip,
 amsmath,amssymb,
 reprint,%
]{revtex4-1}
\usepackage[T1]{fontenc}          
\usepackage[utf8]{inputenc} 
\usepackage{graphicx}
\usepackage{amsmath}
\usepackage{amssymb}
\usepackage{bbm}
\usepackage{dsfont}
\usepackage{newtxmath}
\usepackage{dcolumn}
\usepackage{babel}
\usepackage{color}
\usepackage[normalem]{ulem}

\draft 

\begin{document}


\title{A Rapid Thermal Chemical Vapor Deposition System for Fast Synthesis of Epitaxial Graphene Under Ambient Pressure} 

\author{Shikhar Kumar Gupta}
\affiliation{Department of Condensed Matter Physics and Materials Science, Tata Institute of Fundamental Research, Homi Bhabha Road, Mumbai 400005, India}

\author{Meet Ghelani}
\affiliation{Department of Condensed Matter Physics and Materials Science, Tata Institute of Fundamental Research, Homi Bhabha Road, Mumbai 400005, India}
\affiliation{Department of Physics, The Institute of Science, Homi Bhabha State University, Mumbai 400032, India}

\author{Pragna Datta}
\affiliation{Department of Condensed Matter Physics and Materials Science, Tata Institute of Fundamental Research, Homi Bhabha Road, Mumbai 400005, India}
\affiliation{Department of Physics, The Institute of Science, Homi Bhabha State University, Mumbai 400032, India}

\author{Subhalakshmi Guha}
\affiliation{Department of Condensed Matter Physics and Materials Science, Tata Institute of Fundamental Research, Homi Bhabha Road, Mumbai 400005, India}
\affiliation{Department of Physics, St. Xaviers College, Kolkata 700016, India}

\author{Shivesh Yadav}
\affiliation{Department of Condensed Matter Physics and Materials Science, Tata Institute of Fundamental Research, Homi Bhabha Road, Mumbai 400005, India}

\author{Nilesh Kulkarni}
\affiliation{Department of Condensed Matter Physics and Materials Science, Tata Institute of Fundamental Research, Homi Bhabha Road, Mumbai 400005, India}

\author{Maheshwar Gokhale}
\affiliation{Department of Condensed Matter Physics and Materials Science, Tata Institute of Fundamental Research, Homi Bhabha Road, Mumbai 400005, India}

\author{Bhagyashree Chalke}
\affiliation{Department of Condensed Matter Physics and Materials Science, Tata Institute of Fundamental Research, Homi Bhabha Road, Mumbai 400005, India}

\author{Devendra Buddhikot}
\affiliation{Department of Condensed Matter Physics and Materials Science, Tata Institute of Fundamental Research, Homi Bhabha Road, Mumbai 400005, India}

\author{Naveen Paneri}
\affiliation{Department of Condensed Matter Physics and Materials Science, Tata Institute of Fundamental Research, Homi Bhabha Road, Mumbai 400005, India}
\affiliation{Department of Physics, Indian Institute of Science Education and Research, Pune 411008, India}

\author{Lavudya Devendar}
\affiliation{Department of Condensed Matter Physics and Materials Science, Tata Institute of Fundamental Research, Homi Bhabha Road, Mumbai 400005, India}

\author{Arnab Bhattacharya}
\affiliation{Department of Condensed Matter Physics and Materials Science, Tata Institute of Fundamental Research, Homi Bhabha Road, Mumbai 400005, India}

\author{Shouvik Chatterjee}
\email[Authors to whom correspondence should be addressed: ]{shouvik.chatterjee@tifr.res.in}
\affiliation{Department of Condensed Matter Physics and Materials Science, Tata Institute of Fundamental Research, Homi Bhabha Road, Mumbai 400005, India}



\begin{abstract}

Graphene has emerged as a promising material for next-generation electronic and thermal devices owing to its exceptional charge transport and thermal conductivity. However, high-quality samples are predominantly obtained via mechanical exfoliation from graphite crystals, a process that inherently lacks scalability. Despite extensive efforts toward large-area synthesis, cost-effective approaches for producing high-quality, large-area, single-crystalline graphene with fast turnaround time remain limited. Here, we report the design, fabrication, and performance benchmarking of a rapid thermal chemical vapor deposition (RTCVD) system capable of synthesizing epitaxial monolayer graphene under atmospheric pressure. The entire growth process, from sample loading to unloading, is achieved within $25$ minutes with a temperature ramp rate exceeding $23^\circ\mathrm{C}/s$. Growth at atmospheric pressure eliminates the need for vacuum components, thereby reducing both system complexity and operational costs. The structural and electronic quality of epitaxial graphene is comprehensively characterized using Raman spectroscopy, selected area electron diffraction (SAED), and magnetotransport measurements, which reveal signatures of quantum Hall effect in synthesized graphene samples. Furthermore, we demonstrate van der Waals epitaxial growth of palladium (Pd) thin films on graphene transferred to Si/SiO$_{2}$ substrates, establishing its single-crystalline nature over a large area and its potential as a versatile platform for subsequent heteroepitaxial growth.
 
\end{abstract}


\maketitle 

\section{Introduction}

Graphene, a two-dimensional sheet of carbon atoms arranged in a hexagonal lattice, has attracted immense attention due to its remarkable electronic, thermal, and mechanical properties, and is being extensively investigated for both fundamental studies and technological innovations. Many of these applications critically depend on the availability of large-area, pristine, single-crystalline graphene, which cannot be achieved through mechanical exfoliation of graphite. Furthermore, the exfoliation process lacks scalability, making it unsuitable for industrial or device-level applications. A widely adopted approach to synthesize large-area graphene involves direct growth on catalytic substrates using chemical vapor deposition (CVD). In this process, a transition metal typically serves as a catalyst and is heated to high temperatures (around $1000^\circ\mathrm{C}$). A carbon precursor, most commonly methane, is then introduced. The catalytic surface of the metal promotes methane dissociation, allowing carbon atoms to deposit and form graphene during the subsequent cooling stage~\cite{mechanism}, with the outcome influenced by the carbon solubility of the metal.
The efficiency of this process is largely governed by the interaction between the vacant $d$ orbitals of the transition metal and the $p$ orbitals of carbon~\cite{dpoverlap}.

The heating method employed in the CVD process is a critical determinant of the quality of synthesized graphene. In conventional hot-wall CVD systems, the entire growth tube is uniformly heated, indirectly warming the substrate. Although conceptually straightforward, this configuration introduces several drawbacks including contamination from degradation of quartz or ceramic tubes (e.g. SiO$_{2}$ cluster formation)~\cite{SiO2cont}, enhanced gas-phase reactions leading to amorphous carbon deposition~\cite{Amorphouscarbondeposition}, and slow heating and cooling rates, typically limited to $\sim10^\circ\mathrm{C}/\mathrm{min}$, due to power constraints and thermal-shock concerns, particularly in ceramic tubes. In addition, the high operating temperatures poses safety hazards such as potential hydrogen leakage at elevated temperatures. Cold-wall CVD, on the other hand, offers a superior alternative by enabling localized, substrate-specific heating that minimizes gas-phase reactions and contamination while allowing rapid thermal ramping for both heating and cooling~\cite{fastcoolingofgraphene}. Such fast temperature ramp rates not only improve control over the reaction kinetics but also suppress copper (Cu) evaporation during growth, thereby preserving the Cu(111) surface necessary for high-quality epitaxial graphene formation. Localized heating in a cold-wall CVD can be realized through various techniques, including halogen-lamp irradiation~\cite{HalogenLamp,CommercialRTA}, pulsed-laser heating~\cite{pulselaserdeposition}, plasma-enhanced CVD~\cite{plasmaenhancedcvd}, and RF induction heating~\cite{RFinductionSiC}. However, many of these approaches are technologically complex or depend on expensive commercial infrastructure. Moreover, high-quality graphene growth is typically performed under reduced pressure, as it allows better control over nucleation and helps minimize defect density~\cite{slowgrowthrate}. However, operating under low pressure increases system cost and complexity, making it unsuitable for high-throughput or continuous production. In contrast, atmospheric-pressure CVD offers a more cost-effective and scalable alternative, but often results in thicker and more defective graphene due to enhanced gas-phase reactions and limited surface diffusion~\cite{coldwallcvd}. Therefore, although numerous techniques have been developed to scale up graphene production, most still suffer from one or more drawbacks, such as high cost, slow growth rate, or poor film quality.

To address all these challenges collectively, we have developed a simple halogen-lamp–based rapid thermal chemical vapor deposition (RTCVD) system optimized for the atmospheric-pressure synthesis of epitaxial graphene. The system employs a compact architecture with minimal infrastructure requirements, making it both simple and cost-effective to construct. This design enables fast, reproducible, and scalable graphene growth, producing high-quality, large-area, single-crystalline monolayer films within a total process time of less than 25 minutes. 

We have performed comprehensive investigation of the quality of the synthesized graphene layers using Raman spectroscopy, selected area electron diffraction (SAED), and magnetotransport measurements to benchmark them against existing reports in the literature. Clear signatures of the quantum Hall effect are observed in Hall bar devices fabricated directly on graphene transferred onto Si/SiO$_{2}$ substrates via a wet transfer process without any hBN encapsulation. The measured carrier mobilities at both room and low temperatures rival the highest values reported for graphene transferred onto Si/SiO$_{2}$ using comparable wet transfer techniques\cite{Wet_Transfer_mobility}. Furthermore, we demonstrate van der Waals epitaxial growth of palladium (Pd) thin films on graphene transferred onto Si/SiO$_{2}$ substrates. Our results not only confirm the single-crystalline nature of the synthesized graphene at macroscopic length scales but also highlight its potential as a versatile flexible template for heteroepitaxial integration of diverse material systems. These results establish the efficacy of the RTCVD system as a fast, reliable, and scalable approach for producing large-area, high-quality, single crystalline graphene.

\begin{figure*}[t] 
    \centering
    \includegraphics[scale=1]{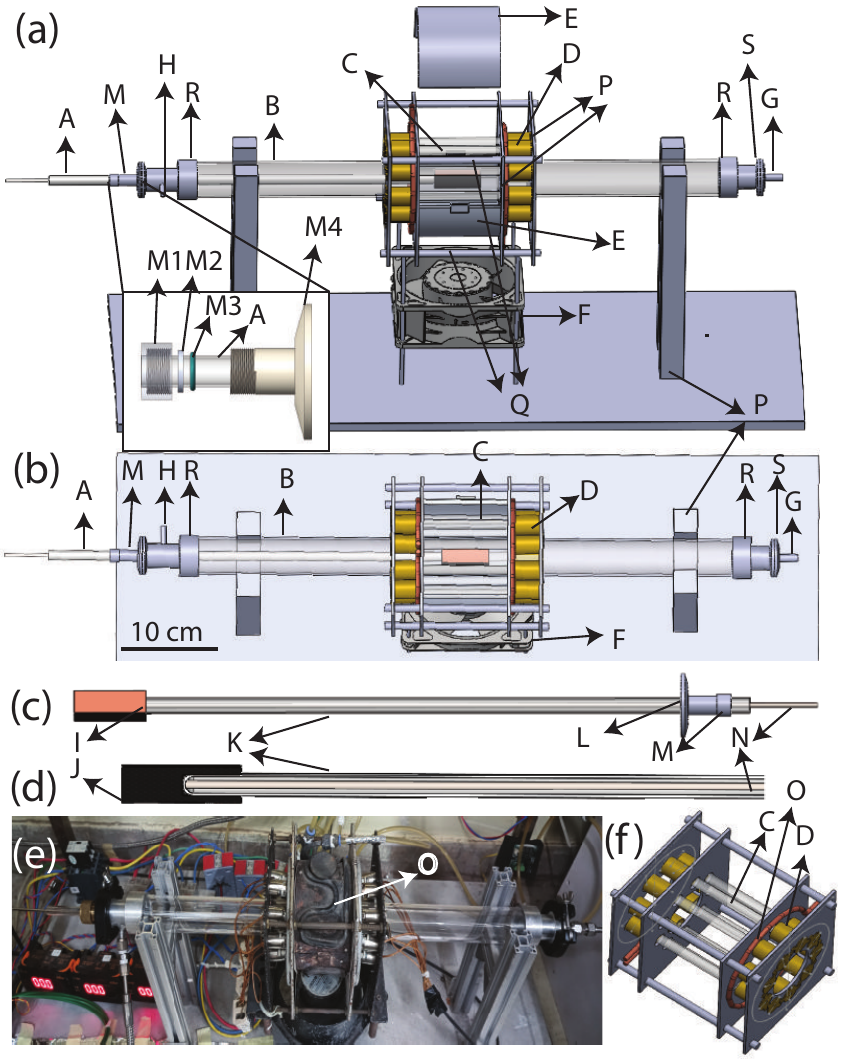}
    \caption{(a) Front view and (b) top view of the RTCVD setup. (c) The sample loader assembly consisting of a smaller quartz tube with the graphite block, along with Cu foil on top (d) Cross-sectional view of the sample loader showing the position of the thermocouple. (e) An image of the RTCVD setup (f) Lamp holder assembly. All the individual components of the RTCVD setup are shown with an arrow: (A) sample loader (B) larger quartz tube (C) halogen Lamp (D) connector (E) housing for Al reflector (F) fan (G) gas inlet (H) gas outlet (I) Cu foil (J) graphite block (K) smaller quartz tube (L) KF40 connector (M) Wilson seal (M1) inner thread (M2) Al ring (M3) viton O-ring (M4) KF connector (N) thermocouple (O) cooling lines connected to chiller. (P) SS plates (Q) screw rods (R) Wilson seal coupler (S) KF40 coupler. } 
    \label{fig:RTA_System}
\end{figure*}

\section{Design and Working Principle}

The heating principle of the RTCVD system relies on radiative energy transfer from halogen lamps that predominantly emits in the infrared region\cite{halogen_spectrum}, to a graphite susceptor. Graphite acts as a gray body in the near-IR region \cite{emissivityofgraphite,emissivityofgraphite2}, efficiently converting radiative power into heat. A Cu substrate, placed directly on top of the graphite block, is heated primarily by conduction. Due to the high infrared reflectivity of metals, the use of graphite susceptor is essential for efficient power coupling. Furthermore, since growth happens within a quartz tube, which is transparent to the IR radiation, the whole apparatus acts as a cold-wall CVD, where only the graphite susceptor and the sample placed on top of it gets heated. Once the required substrate temperature is reached, the process gases are introduced for graphene growth.

The RTCVD system, shown in Fig.~\ref{fig:RTA_System}, consists of nine halogen lamps (component (C) in Fig.~\ref{fig:RTA_System}) arranged in a circular geometry and completely enclosed by 0.2 mm thick aluminium (Al) sheet reflectors. Each lamp is rated at 220V, 1 kW, and the nine lamps are divided into three independently controlled sets, each powered by a separate single-phase supply. Within each set, the lamps are connected in parallel, and the current in each phase is monitored using an AC ammeter. The Al reflectors are riveted to the inner surfaces of two semi-circular, water-cooled stainless steel (SS304) shields (component (E) in Fig.~\ref{fig:RTA_System}). Active cooling of the reflector assembly is achieved by circulating chilled water at $12^\circ$C at a flow rate of $1.7~\mathrm{l/min}$. An auxiliary cooling fan located at the base (component (F) in Fig.~\ref{fig:RTA_System}) further enhances heat dissipation from the external components. The Al sheets (type W13C1) possess a high-quality surface finish with an average roughness of 0.01–0.02 $\mathrm{\mu m}$ and an optical reflectivity of 85$\%$. Their infrared reflectivity exceeds 90$\%$~\cite{ReflectivityofAl}, ensuring efficient heat confinement. A 600 mm long quartz tube (component (B) in Fig.~\ref{fig:RTA_System}) passes coaxially through the lamp assembly, providing an infrared-transparent barrier that isolates the process environment while allowing nearly all incident radiation to reach the graphite susceptor. A high-density graphite block (50 mm $\times$ 15 mm $\times$ 15 mm) is positioned within this tube and serves as the primary heat absorber. One end of the graphite block contains a circular cavity extending to its center and snugly fits onto a smaller quartz tube (component (A) in Fig.~\ref{fig:RTA_System}), concentric with the larger one. The smaller tube is closed at the end supporting the graphite block and open at the other end, allowing insertion of a thermocouple (component (N) in Fig.~\ref{fig:RTA_System}). A Wilson seal near the open end (see Fig.~\ref{fig:RTA_System}(a)) isolates the smaller tube from the larger quartz tube. The thermocouple tip remains in direct contact with the sealed end, on which the graphite susceptor is push-fit, and is positioned at the center of the block (see Fig.~\ref{fig:RTA_System}(d)), ensuring accurate temperature measurement of the sample placed above it. The high thermal conductivity of graphite ensures uniform temperature distribution across the sample surface. All nine lamps are operated simultaneously under PID control, using thermocouple feedback with parameters optimized near the growth condition. To suppress carbon contamination from the graphite block, the top surface of the susceptor is covered with a commercial high-purity, polycrystalline Cu foil (Fig.~\ref{fig:RTA_System}(c)), on which the sample is directly placed. Alternatively, a SiC/TaC-coated graphite block may be employed to achieve similar protection.

\begin{figure}
\centering
 \includegraphics[width=1\linewidth]{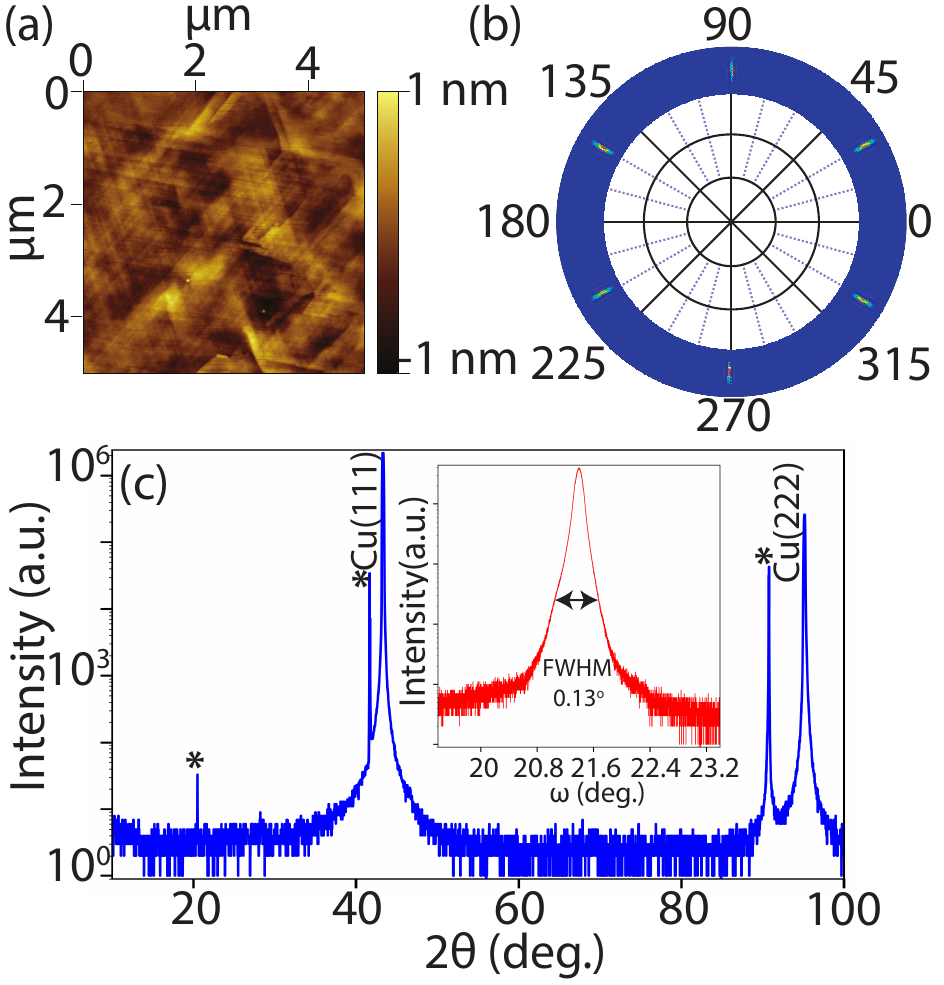}
\caption{(a) AFM image of 1 $\mu$m thick Cu film before graphene growth. RMS roughness is 1.6 nm over 5 x 5 $\mu m^{2}$. (b) Pole figure of Cu (002) plane showing 6 fold symmetry due to the presence of twins on the Cu (111) surface (c) Out-of-plane $\theta-2\theta$ scan of Cu establishing the presence of only ($lll$) diffraction planes. Substrate peaks are marked by asterisks. (b) and (c) establish the epitaxial nature of Cu thin films. Inset show rocking curve of Cu (111) diffraction peak with a FWHM of 0.13$^\circ$.
\label{Cu_charac}}%
\end{figure}

\begin{figure}
\centering
 \includegraphics[width=1\linewidth]{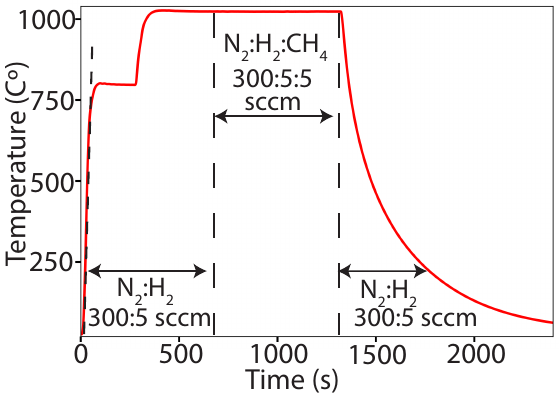}
\caption{Temperature profile of graphene growth in RTCVD\label{temp_profile}}%
\end{figure}

The entire lamp assembly is mounted using two aluminium plates (component (P) in Fig.~\ref{fig:RTA_System}) positioned on either side and supported by four threaded rods (component (Q) in Fig.~\ref{fig:RTA_System}). Each of the four plates is machined with nine concentric holes to accommodate the lamps and their holders, along with a central aperture (50 mm in diameter) for the quartz tube. The inner pair of plates holds the lamps in place, while the outer pair supports the electrical connectors (component (D) in Fig.~\ref{fig:RTA_System}). The complete assembly - comprising the four plates and threaded rods - is mounted on a cooling fan, as described earlier, to aid in heat dissipation. The quartz process tube is additionally supported at both ends and interfaced with the gas lines through aluminium couplers secured with Wilson seals (component (R) in Fig.~\ref{fig:RTA_System}). Samples are introduced into the larger quartz tube through the smaller coaxial quartz tube (see Fig.~\ref{fig:RTA_System}(c)) from the exhaust side. Because the exhaust end of the larger tube is sealed by the loading assembly, an auxiliary outlet line is incorporated via a KF40 coupler (component (H) in Fig.~\ref{fig:RTA_System}) to maintain gas flow. The circular configuration of the lamp array provides a compact footprint and facilitates straightforward integration with standard gas-handling and pressure-control systems. The system achieves a heating rate exceeding $\mathrm{23^\circ C/s}$ (see Fig.~\ref{temp_profile}) and cools from growth temperature to room temperature within 15 minutes, which significantly contributes to the quality of the synthesized graphene\cite{coldwallcvd,fastcoolingofgraphene}. The overall power efficiency of the setup is estimated to be approximately 10$\%$.

In this work, epitaxial graphene is synthesized in the RTCVD system directly on Cu(111) thin films, which are epitaxially grown on c-plane sapphire substrates by magnetron sputtering. Among various transition metal elements used for catalytic cracking of methane, Cu stands out due to its low cost and extremely low bulk solubility for carbon~\cite{carbonsolubility}. Moreover, the Cu(111) surface offers an epitaxial template for graphene growth, with a lattice mismatch of only 3.6$\%$~\cite{epitaxialCu,epitaxialgraphene}. These attributes make the Cu(111) surface an ideal substrate for the high-quality synthesis of epitaxial graphene, as employed in this work. 

The issue of enhanced gas-phase reactions under atmospheric pressure is mitigated by employing a low methane concentration, which limits dehydrogenation and reduces the number of carbon species available for deposition. Since the RTCVD system effectively works as a cold-wall CVD reactor, graphene growth is governed primarily by surface-mediated reactions. Once the Cu surface becomes completely covered with a graphene layer, the limited availability of reactive carbon species in the gas phase suppresses further nucleation and growth, leading to self-limited monolayer formation. Moreover, the absence of wall heating eliminates potential silicon contamination arising from quartz tube degradation~\cite{Amorphouscarbondeposition}, while the suppression of gas-phase reactions minimizes memory effects between growth runs, ensuring long-term reproducibility.

Although atmospheric operation generally increases nucleation density, the epitaxial alignment of the Cu(111) surface promotes rapid domain coalescence, yielding a continuous, uniform graphene monolayer. Furthermore, operating under atmospheric pressure offers several practical and material advantages. First, it suppresses copper evaporation at elevated temperatures - an essential consideration for sputtered Cu(111) thin films, where achieving several-micron thicknesses is often impractical. This suppression minimizes copper loss, and together with potential reusability of the c-plane sapphire substrates improves the economic viability of the growth process. Second, atmospheric-pressure operation reduces impurity incorporation from the graphite susceptor, which absorbs infrared radiation and locally heats the substrate during growth. In addition, we demonstrate that nitrogen (N$_{2}$) can effectively replace argon (Ar) as the process gas, further lowering operational costs. As N$_{2}$ remains chemically inert at $\sim1000^\circ\mathrm{C}$, no N$_{2}$ incorporation into the graphene lattice is observed~\cite{N2dopinggrowth}. Combined with the fast heating and cooling capabilities of the halogen-lamp RTCVD system these advantages enable efficient exploration of growth parameters with a rapid turn around time, while providing a low-cost, scalable pathway for synthesizing high-quality, single-crystalline, large-area graphene.

\section{Growth Condition and Characterization}

 High-quality Cu(111) thin films were first deposited using DC magnetron sputtering, the details of which can be found in the Supplementary Information\cite{suppl}. The epitaxial nature and (111) out-of-plane orientation of the Cu films were confirmed by x-ray diffraction (XRD), as shown in Fig.~\ref{Cu_charac}(b,c). Owing to the lower in-plane symmetry of the Cu(111) surface relative to that of c-plane sapphire, the films exhibit twinning, consistent with previous reports\cite{copperepitaxysapp}. However, these twin boundaries do not significantly degrade graphene quality, as graphene can grow seamlessly across them without introducing line defects~\cite{grapheneovertwin}. In this work, we demonstrate that high-quality epitaxial graphene can indeed be synthesized on such twinned Cu(111) surfaces~\cite{epitaxialCu} using the RTCVD system. The temperature profile used for graphene growth in the RTCVD system is shown in Fig.~\ref{temp_profile}. Growth was conducted under a continuous flow of $N_2$ (300 sccm) and $H_2$ (5 sccm), with an initial annealing step at $800^\circ\mathrm{C}$. The final growth step was performed at $1030^\circ\mathrm{C}$ with a $CH_4$ flow of 5 sccm for approximately 10 minutes. The entire process was carried out at atmospheric pressure. Synthesized graphene was transferred onto Si/SiO$_2$ substrates using a standard wet-etching method. The sample was first spin-coated with PMMA (A4 495) at 2000 rpm and baked at $120^\circ\mathrm{C}$ for 5 minutes. The underlying Cu film was then etched in a 0.3 M aqueous solution of $(NH_4)_2S_2O_8$, after which the PMMA-supported graphene layer was scooped onto the Si/SiO$_2$ substrate. Finally, PMMA was removed by dissolving it in acetone.

\begin{figure}
\centering
  \includegraphics[width=1\linewidth]{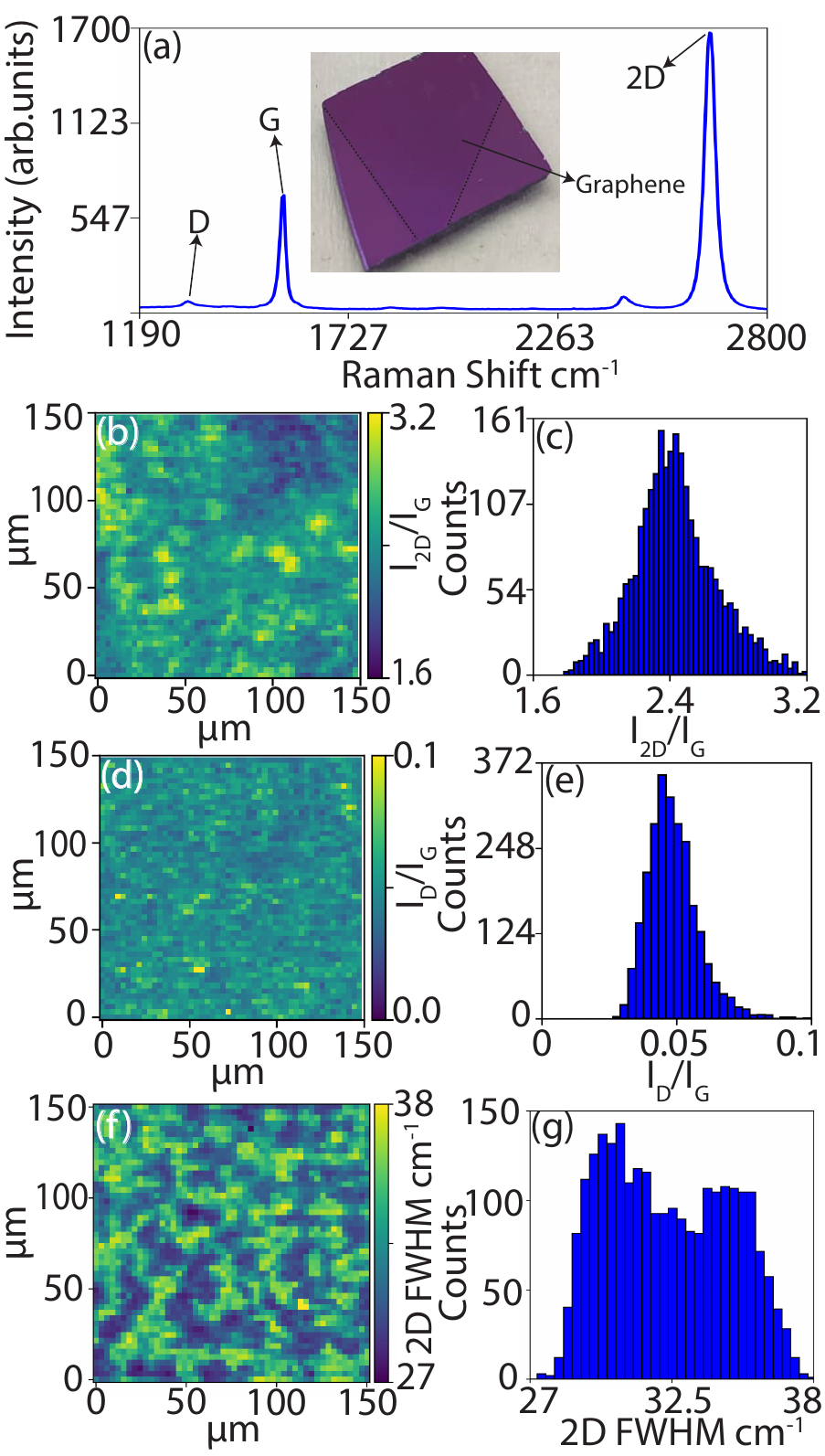}
\caption{(a) Raman spectrum obtained after averaging over a $150\times150~\mu\mathrm{m}^2$ area using a $50\times50$ grid, taken from graphene transferred to Si/SiO$_{2}$ substrate, showing the characteristic $G$, $2D$, and the defect $D$ peaks. Inset shows an optical image of the graphene transferred to Si/SiO$_{2}$ substrate on which Raman spectroscopy was done. Raman maps showing the distribution of (b) intensity ratio of the $2D$ and $G$ peaks - $I_{2D}/I_{G}$ (d) intensity ratio of the $D$ and $G$ peaks - $I_{D}/I_{G}$ and (f) FWHM of the $2D$ peak over a $150 \times 150~\mu\mathrm{m}^2$ field of view. Corresponding histogram plots are shown in (c), (e), and (g), respectively.\label{Graphene_map}}%
\end{figure}

Raman spectroscopy was performed on graphene transferred onto Si/SiO$_{2}$ substrates to evaluate the uniformity and thickness of the synthesized films over large areas. The optical image in Fig.~\ref{Graphene_map}(a) confirms the continuous, crack-free transfer of the graphene layer. Raman mapping was carried out over a $150\times150~\mu\mathrm{m}^2$ area using a $50\times50$ grid with a 20$\times$ objective on a WITec alpha300R Raman system. The spatial maps of the $G$, $2D$, and defect-related $D$ peaks, located at 1585, 2690, and 1350~cm$^{-1}$, respectively, were analyzed to extract the intensity ratios $I_{2D}/I_G$ and $I_D/I_G$. The average values obtained were $I_{2D}/I_G = 2.43\pm0.24$ (Fig.~\ref{Graphene_map}(b)) and $I_D/I_G = 0.05\pm0.01$ (Fig.~\ref{Graphene_map}(d)), confirming that the synthesized graphene is predominantly monolayer and exhibits very low defect density\cite{Graphene_Spectrum}. The full width at half maximum (FWHM) of the $2D$ peak was measured to be $32.34\pm2.43~\mathrm{cm^{-1}}$, among the narrowest reported for graphene transferred using similar wet-etching technique~\cite{FWHM_2dpeak_1,FWHM_2dpeak_2}, as employed in this work. Comparable Raman characteristics were reproducibly obtained across more than 300 independent growth runs using the RTCVD system, underscoring the reliability and consistency of our synthesis method\cite{suppl}.

\begin{figure}
\centering
 \includegraphics[width=1\linewidth]{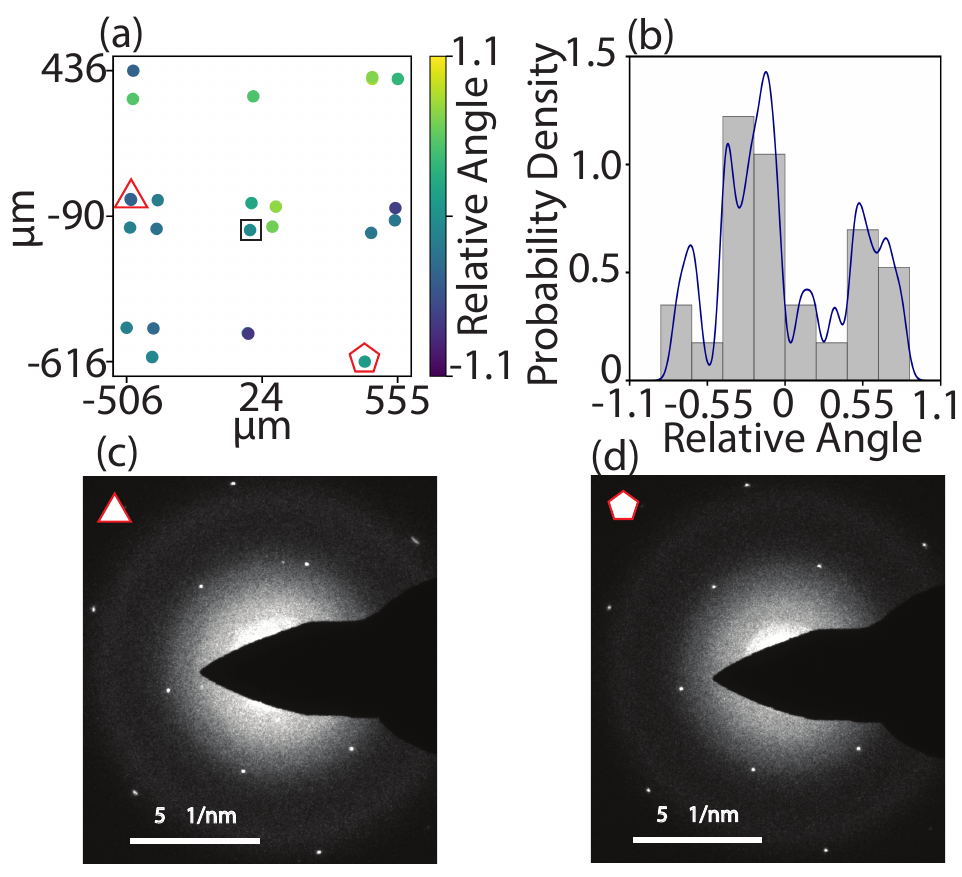}
\caption{(a) Regions where SAED measurements were performed, the square represents the position where the reference image was taken, as discussed in the main text. (b) Distribution of the relative orientation angles calculated from SAED images at different regions with respect to the reference image. Representative SAED images taken at two different regions marked with (c) a triangle and (d) a pentagon, respectively, in (a). \label{epitaxy_raman}}%
\end{figure}

Having established the synthesis of high-quality, continuous, large-area monolayer graphene, we next examine its epitaxial nature. Selected area electron diffraction (SAED) measurements were performed using a TECNAI 20; 200 kV TEM with LaB6 filament with 40 $\mu$m aperture on graphene transferred onto SiN TEM grids via the same wet-etching technique described earlier. The SAED patterns, shown in Fig.~\ref{epitaxy_raman}, exhibit sharp, six-fold symmetric diffraction spots, confirming the single-crystalline nature of the synthesized graphene. To assess the in-plane rotational alignment across large areas, 24 SAED patterns were acquired from multiple locations over a $1\times1~\mathrm{mm^2}$ region. Each diffraction pattern was referenced to that of a randomly chosen location (marked by a square in Fig.~\ref{epitaxy_raman}(a)). The resulting angular distribution, shown in Fig.~\ref{epitaxy_raman}(b), reveals an average spread of approximately $0.46^\circ$, demonstrating excellent epitaxial alignment and confirming the single-crystalline nature of the graphene film over macroscopic length scales. Minor angular deviations likely originate from small cracks introduced during the transfer process, which can locally create or relax strain and induce slight rotational offsets in the graphene lattice.

\begin{figure}
\centering
 \includegraphics[width=1\linewidth]{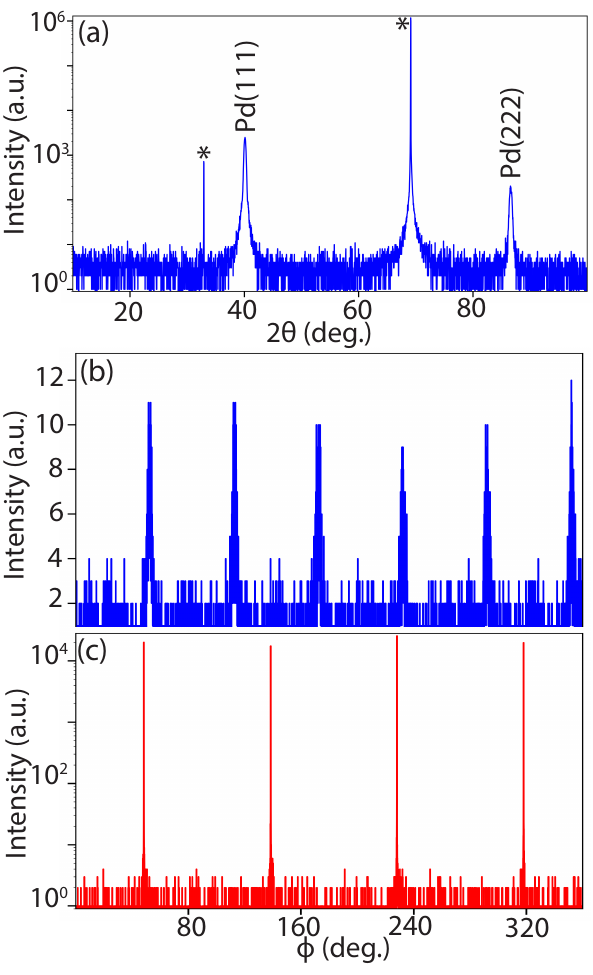}
\caption{(a) Out-of-plane $\theta$-2$\theta$ XRD scan establishing the presence of only Pd($lll$) diffraction peaks. Diffraction peaks from Si/SiO$_{2}$ substrate are marked by asterisks. Azimuthal $\phi$ scan of asymmetric (b) Pd(022) and (c) Si(022)\label{epitaxy_pd}}%
\end{figure}

Additional confirmation of the single-crystalline nature of the synthesized graphene over large area is provided by the demonstration of van der Waals epitaxy of palladium (Pd) thin films on graphene transferred to Si/SiO$_2$ substrates. Graphene was transferred using the same wet-etching procedure described earlier, followed by vacuum annealing at $600^\circ$C for 2~hours under a flow of 30~sccm H$_2$ and 100~sccm N$_2$ to remove residual PMMA. Subsequently, the sample was loaded into a molecular-beam epitaxy (MBE) chamber, where a 10~nm thick Pd film was deposited using e-beam evaporation at a rate of $\approx$0.17~nm/min. Pd nucleation was initiated at $160^\circ$C, followed by deposition and post-growth annealing for 30 minutes, both performed at $500^\circ$C sample temperature. XRD measurements confirm that the Pd films are single-crystalline with a (111) out-of-plane orientation, shown in Fig.~\ref{epitaxy_pd}(a). The azimuthal $\phi$ scan of the asymmetric Pd(022) reflection exhibits six-fold symmetry (see Fig.~\ref{epitaxy_pd}(b)), indicating that the Pd films are single-crystalline but contain twin domains - consistent with epitaxial alignment on the six-fold symmetric single-crystalline graphene lattice. It is noteworthy that the underlying Si/SiO$_2$ substrate consists of an amorphous $\sim$285~nm SiO$_2$ layer on top of Si(001), a four-fold symmetric surface (see Fig.~\ref{epitaxy_pd}(c)). Therefore, the observed epitaxial alignment of Pd(111) is mediated solely by van der Waals epitaxy on graphene, providing direct evidence that the transferred graphene is single-crystalline over macroscopic length scales. This, in turn, validates the epitaxial growth of high-quality graphene on Cu(111) achieved using the RTCVD system.

\begin{figure}
\centering
 \includegraphics[width=1\linewidth]{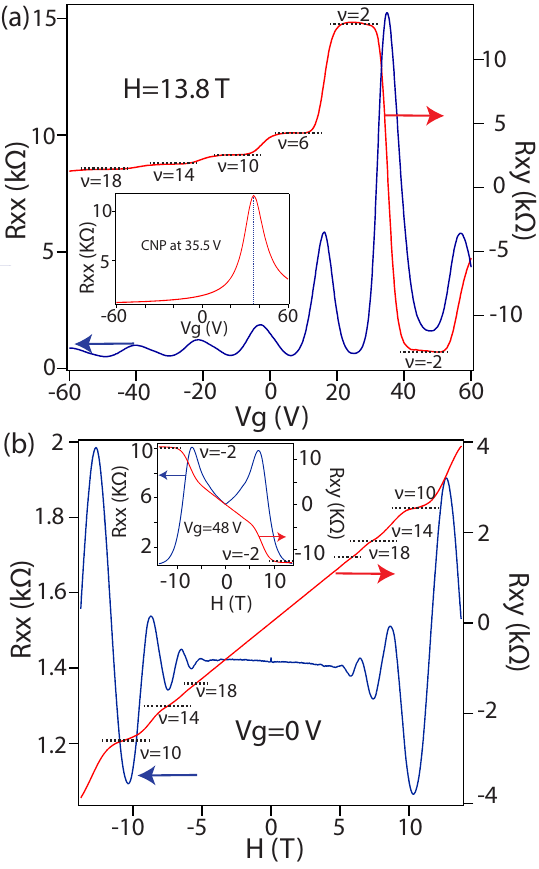}
\caption{(a) Longitudinal (R$_{xx}$) and Hall (R$_{xy}$) resistance as a function of gate voltage at 2 K and a magnetic field of 13.8 T. Inset shows R$_{xx}$ as a function of gate voltage at 2 K and zero magnetic field with the charge neutrality point (CNP) at 35.5 V. (b) R$_{xx}$ and R$_{xy}$ as a function of magnetic field at 2 K at zero gate voltage. Observation of positive slope in (b) indicates the presence of hole-like carriers. Inset shows R$_{xx}$ and R$_{xy}$ as a function of magnetic field at 2 K at a gate voltage of 48 V indicating the presence of electron-like carriers.
\label{Halldata}}%
\end{figure}
Finally, to probe the electronic properties of the synthesized graphene, magnetotransport measurements were carried out on samples transferred onto Si/SiO$_2$ substrates using the same wet-transfer method described earlier. Hall bar devices with channel dimensions of $\mathrm{40~\mu m}$ $\times$ $\mathrm{10~\mu m}$ were fabricated using standard electron-beam lithography, details of which can be found in the Supplementary Information\cite{suppl}. Electrical transport measurements were performed in a commercial Physical Property Measurement System (PPMS) equipped with a 13.8~T superconducting magnet and a base temperature of 2~K. The back-gate voltage was applied through the p-doped Si(001) substrate, with the 285~nm thick SiO$_2$ layer serving as the gate dielectric. The gate-voltage dependence of resistance at zero magnetic field, shown in the inset of Fig.~\ref{Halldata}(a), exhibits the characteristic ambipolar field effect with a charge-neutrality (Dirac) point near +35.5~V and a resistance modulation ratio of $\sim$17. The positive shift of the Dirac point indicates unintentional p-type doping, most likely originating from polymer residues introduced during the transfer and lithography processes - an effect commonly observed in graphene devices on Si/SiO$_2$ prepared via wet transfer\cite{P_doping_PMMA}.

Hall resistance ($R_{xy}$) was measured both as a function of gate voltage at a fixed magnetic field and as a function of magnetic field at different fixed gate voltages, as shown in Fig.~\ref{Halldata}. The data exhibit clear signatures of half-integer quantum Hall plateaus, $R_{xy} = h/(\nu e^{2})$ with $\nu = 4n + 2$ ($n \in \mathbb{I}$), a definitive hallmark of high-quality monolayer graphene. Distinct plateaus corresponding to filling factors $\nu = 2, 6, 10, 14,$ and $18$ on the hole side and $\nu = -2$ on the electron side were observed at 2~K under a maximum magnetic field of 13.8~T, with the applied back-gate voltage swept between $\pm$60~V. The extracted Hall mobilities for electron and hole carriers are 6524~cm$^{2}$/V·s and 6978~cm$^{2}$/V·s, respectively, at 2~K, and 4853~cm$^{2}$/V·s and 6373~cm$^{2}$/V·s, respectively, at 300~K. The observation of well-resolved half-integer quantum Hall plateaus together with high carrier mobilities confirms the exceptional crystalline and electronic quality of the graphene synthesized using the RTCVD method described here. Notably, these transport properties are among the best reported for graphene devices fabricated on Si/SiO$_2$ substrates without employing hBN encapsulation or optimized van der Waals–mediated using oxhydryl groups-containing volatile molecules (OVM) transfer techniques~\cite{OVM_transfer,RollTORoll_mobility,Ulta-Clean_mobility,WaferScale_mobility,ControlledTwinfree_mobility,LargeArea_Mobility}. Further improvements in device processing, including hBN encapsulation and optimized transfer procedures, are expected to enable epitaxial graphene synthesized by RTCVD to reach electronic performance comparable to that of mechanically exfoliated graphene from graphite single crystals.

\section{Conclusion}

In summary, we have demonstrated the design and construction of a low-cost, halogen-lamp-based RTCVD system capable of synthesizing large-area, high-quality epitaxial graphene under atmospheric pressure. The resulting single-crystalline graphene exhibits the half-integer quantum Hall effect and functions as an epitaxial template for the heteroepitaxial integration of diverse material systems via van der Waals epitaxy, underscoring the effectiveness of the RTCVD approach. A practical graphene synthesis technique must be fast, cost-effective, simple, reliable, and reproducible while yielding high-quality films. The RTA-based method developed here fulfills all these requirements, establishing it as a promising route for scalable production of high-quality, single-crystalline, large-area graphene.

\section*{Acknowledgments}

We acknowledge the Department of Science and Technology (DST), SERB grant SRG/2021/000414 and Department of Atomic Energy (DAE), 12-R$\&$D-TFR-5.10-0100, of the Government of India for support. 


\section*{Author Contributions}

S.C. and S.K.G. conceived the project and planned the experiments. S.C. and S.K.G., with assistance from Ma.G. and A.B., designed the experimental apparatus. S.K.G., with assistance from N.K., Ma.G, M.G., and D.B., under the guidance of S.C. and A.B., built the experimental setup. S.K.G., with assistance from M.G., P.D. and Su.G., fabricated epitaxial graphene films, performed Raman spectroscopy, and optimized the wet transfer process of graphene on Si/SiO$_{2}$ substrates. B.C. performed SAED measurements of epitaxial graphene, which were transferred on SiN TEM grids by S.K.G. Synthesis of epitaxial Pd on graphene was performed by S.K.G. and N.P. L.D. helped with x-ray diffraction and AFM measurements. S.K.G. and S.Y. fabricated devices and performed magnetotransport measurements. S.C. was responsible for the overall execution of the project. All authors discussed the results and commented on the manuscript.

\section*{Data Availability}

The data that supports the findings of this study are available from the corresponding author upon reasonable request.

\section*{Competing Interests}
The authors declare no competing interests.

\section*{References}



\end{document}